# Black Hole Radiation and Volume Statistical Entropy

**Mario Rabinowitz**[1]

The simplest possible equation for Hawking radiation $P_{SH} = \frac{G\rho\hbar}{90}$, and other black hole radiated power is derived in terms of black hole density, $\rho$. Black hole density also leads to the simplest possible model of a gas of elementary constituents confined inside a gravitational bottle of Schwarzchild radius at tremendous pressure, which yields identically the same functional dependence as the traditional black hole entropy $S_{bh} \propto (kAc^3)/\hbar G$. Variations of $S_{bh}$ can be obtained which depend on the occupancy of phase space cells. A relation is derived between the constituent momenta and the black hole radius $R_H$, $p = \left(\frac{3}{2\pi}\right)\frac{\hbar}{R_H}$, which is similar to the Compton wavelength relation.

## 1. INTRODUCTION

The object of this paper is to gain an insight into black holes. This is first done by finding the simplest possible equation for black hole radiation. Then we shall find the simplest possible model which can give some comprehension as to why black hole entropy appears to be related only to its surface area. This attempt at understanding black hole entropy is done in the same spirit as was the paper on Classical Tunneling (Cohn and Rabinowitz, 1990) which showed how far a simple classical model can go to illume the phenomenon of quantum tunneling.

[1]Armor Research, 715 Lakemead Way; Redwood City, California 94062-3922

The strategy here will be to find the number of elementary constituents, N, which fill a black hole, simply modeled as a gravitational bottle. The constitutents are unspecified and may be gravitons or more exotic entities such as branes. We shall find that they have a mass that is inversely proportional to the horizon radius of the black hole. Once we have N, statistical mechanics will be employed to determine their entropy, although it is not clear that it is applicable to black holes. And even if it is applicable, it is not obvious how to apply it. We can see this by briefly considering the applicability and limitations of Liouville's theorem with respect to black hole thermodynamics, since it is so central to statistical mechanics. In turn, statistical mechanics represents a foundation for thermodynamics.

First of all Liouville's theorem applies only to a non-dissipative system in which energy (KE + PE) is conserved. To a good approximation energy is conserved for a large black hole as it hardly radiates, but this is not a good approximation for a small black hole due to Hawking radiation. The system of any size black hole and Hawking radiation conserves mass-energy. However, this is not conserved for the black hole itself.

Second and perhaps even more importantly, correlation with thermodynamics depends subtly on the ergodic hypothesis, since it is assumed that the system can indeed move from one region of phase space to another based upon the equations of motion, consistent with conservation of energy. The ergodic hypothesis implies that every state of the system can be reached directly or indirectly from every other state. This means that if the energy of the system is determined within a range $\Delta E$, the probability of finding the system in a certain state compatible with that energy is the same for each state.

Third and most importantly in not only the ergodic hypothesis, but as the basis of statistical mechanics, the time average over the evolution of the system is replaced by the average over the different states. Black hole thermodynamics in general, and black hole entropy in particular does not clearly specify what the states are. Nor is the validity of doing this self-evident in general relativity, because of the elasticity of space-time. How is it applied to the surface of a black hole where time appears to stand still from the perspective of a distant observer?

Although the simple analysis that will be presented (starting with Sec. 3) may seem to also skirt these issues, it circumvents them much less than the area entropy algorithm as well as indicating why it works so well. At least partial success is attained in obtaining an expression for entropy that agrees remarkably well with the black hole entropy functional dependence that was obtained by Bekenstein (1972, 1973, 1974).

It is noteworthy that the prime critic of Bekenstein's 1972 conception was Hawking, who three years later (Hawking,1975) embraced the concept and found the constant of proportionality to be 1/4. Hawking's main criticism had been that if black holes have entropy, they must have temperature, and if they have temperature they must radiate; and everyone knows that black holes can't radiate. He withdrew this criticism when he realized that black holes can radiate.

## 2. SIMPLEST EQUATIONS FOR BLACK HOLE RADIATION

Hawking (1974, 1975) introduced what is now called Hawking radiation as the effective black body radiation from a black hole in terms of the 4th power of the black hole temperature and the Stefan-Boltzmann constant. In terms of fundamental parameters the power radiated from a black hole of mass M is given by (Rabinowitz, 1999):

$$P_{SH} = \frac{\hbar c^6}{960 \pi G^2 M^2}. \tag{1}$$

The mass density of a black hole is

$$\rho = \frac{M}{\left(\frac{4\pi}{3}\right) R_H^3} = \frac{M}{\left(\frac{4\pi}{3}\right)\left(\frac{2GM}{c^2}\right)^3} = \frac{3c^6}{32 \pi G^3 M^2}, \tag{2}$$

where $R_H$ is the horizon (Schwarzchild) radius of the black hole.

Combining Eqs. (1) and (2) yields

$$P_{SH} = \frac{G \rho \hbar}{90}, \tag{3}$$

which is the simplest possible expression for the power radiated from a black hole. It is also the most intuitive in saying that the radiated power is proportional to the black hole density, $\rho$. When the density is high as it is for small black holes, the power is high. When the density is low as it is for large black holes, the power is low.

Other radiation mechanisms such as gravitational tunneling radiation (GTR) generalized to n-space, and higher dimensional entropy are explored in Rabinowitz (2001a, b). If we combine the GTR power with Eq. (2) we obtain

$$P_R = \frac{\hbar c^6}{16 \pi G^2 M^2} \frac{\langle e^{-2\Delta\gamma} \rangle}{M^2} = \frac{2 G \rho \hbar \langle e^{-2\Delta\gamma} \rangle}{3}, \tag{4}$$

where $\langle e^{-2\Delta\gamma} \rangle$ is the black hole transmission probability.

Equations (3) and (4) are intuitive in saying that the radiated power increases as the density of the black hole increases.

### 3. UTILIZING THE IDEAL GAS LAW INSIDE A BLACK HOLE

General relativity (GR) is very non-linear inside the black hole horizon where time and space exchange roles. Even Einstein stated his concerns about black holes, According to GR, after a sufficiently long time a black hole should be empty inside the horizon, except for a theoretical but not a physical singularity at its center. To gain a new insight into the entropy of black holes, let us model a black hole of mass M as a spherical gravitational bottle filled with an ideal free gas of N elementary constituents, of average mass m. This model of an ideal gas inside a black hole leads to an equation that agrees remarkably well with standard black hole entropy. At this early stage, it should be judged by its results rather than by our judgments about what is inside a black hole.

We start with the ideal gas law in terms of the mass density $\rho$ of the black hole

$$P = \left(\frac{N}{V}\right)kT = \left(\frac{\rho}{m}\right)kT, \tag{5}$$

where P is the pressure, $V = \left(\frac{4\pi}{3}\right)R_H^3$ is the volume, the Boltzmann constant $k = 1.38 \times 10^{-23}$ J/K, and T is the temperature.

For the black hole temperature we use (Hawking, 1975):

$$T = \left[\frac{\hbar c^3}{4\pi kG}\right]\frac{1}{M}. \tag{6}$$

Combining Eqs. (2) and (6) we have

$$T = \left[\frac{\hbar c^3}{4\pi kG}\right]\frac{1}{M} = \frac{\hbar}{k}\sqrt{\frac{2\rho G}{3\pi}}. \tag{7}$$

Combining Eqs. (5) and (7), we obtain the pressure inside the black hole

$$P = \left(\frac{\rho}{m}\right)k\left[\frac{\hbar}{k}\sqrt{\frac{2\rho G}{3\pi}}\right] = \frac{\hbar}{m}\left(\frac{2G}{3\pi}\right)^{1/2}\rho^{3/2}, \tag{8}$$

containing N = M/m constituents. From kinetic theory, the pressure is also

$$P \approx \tfrac{1}{3}(N/V)\left[\overline{mv^2}\right] \approx \tfrac{1}{3}\rho c^2. \tag{9}$$

The approximation $v \approx c$ can be justified if the constituents have zero rest mass; or if we are dealing with little black holes that are at such high temperature that that $v \approx c$ even for constituents with rest mass. The question of the consituent rest mass will be explored in Sec. 5. Equating eqs. (8) and (9), and solving for the mass density $\rho$:

$$\rho = \left(\frac{3\pi}{2G}\right)\left(\frac{m^2 c^4}{9\hbar^2}\right) = \left(\frac{N}{V}\right)m. \tag{10}$$

Combining Eqs.(2) and (10) we find the number of constituents inside the black hole:

$$N = \left(\frac{3}{4\pi}\right)\left(\frac{m_P}{m}\right)^2 = \left(\frac{4\pi}{3}\right)\left(\frac{M}{m_P}\right)^2, \tag{11}$$

where the Planck mass $m_P = (\hbar c / G)^{1/2} = 2.18 \times 10^{-8}$ kg.    If an estimate is made for the collision cross section, the collision frequency of the constituents can be calculated from their number density $\rho / m$ as given by Eq. (10)/m. From the collision frequency one can estimate the time for the black hole to reach equilibrium after an interaction with matter or another black hole.

### 4. STATISTICAL MECHANICS VOLUME CONTRIBUTION TO ENTROPY

The standard Boltzmann statistical mechanical entropy of a system of N constituents with $N_s$ distinct states is, for large N:

$$S_{bh} = k \ell n N_s \approx kN. \tag{12}$$

The approximation $\ell n N_s \approx N$ is commonly made in the scientific literature and in textbooks (Mayer and Mayer, 1940). In 1896, Boltzmann was the first to interpret entropy in terms of the total number of [quantum] states available or accessible to a system. According to Sommerfeld (1952) it was Planck, in 1906, that cast Boltzmann's principle in the form of Eq. (12).

Substituting Eq. (11) into Eq. (12)

$$S_{bh} \approx kN = k\left(\frac{4\pi}{3}\right)\left(\frac{M}{m_P}\right)^2 = \frac{kAc^3}{12\hbar G}. \tag{13}$$

Relating Eq. (13) to the Bekenstein (1974) black hole entropy

$$S_{bh} = \frac{1}{3}S_{Bek} = \frac{1}{3}\left[\frac{kAc^3}{4\hbar G}\right]. \tag{14}$$

### 5. CONSTITUENT MASS

By means of Eq. (11), we obtain for the constituent mass

$$m = \left(\frac{3}{4\pi}\right) m_P^2 \left(\frac{2G}{R_H c^2}\right) = \left(\frac{3}{4\pi}\right) \left[\frac{\hbar c}{G}\right] \left(\frac{2G}{R_H c^2}\right) = \left(\frac{3}{2\pi}\right) \frac{\hbar}{c R_H}, \qquad (15)$$

If we can take the limit $R_H \to \infty$, then $m \to 0$, and we can conclude that the constituents have zero rest mass. If $R_H$ is limited to the radius of the universe or some smaller dimension, then the constituents may have a non-zero, but extremely small rest mass.

If we take the momentum $p = mc$ for a zero rest mass constituent, Eq. (15) takes the form of the reduced Compton wavelength of the constituent:

$$\lambdabar = \frac{\hbar}{p} = \frac{\hbar}{mc} = \frac{\hbar}{\left[\frac{3}{2\pi}\left(\frac{\hbar}{R_H}\right)\right]} = \frac{2\pi}{3} R_H \approx 2 R_H. \qquad (16)$$

## 6. DISCUSSION

In the prevailing point of view, it is not clear what distinct black hole states are being counted by $N_s$ in the expression $S_{bh} = k \ln N_s$. A further problem is that since entropy and temperature are statistical quantities dealing with many bodies, what does it mean to speak of them with respect to a black hole viewed as a single body, which is all that the Schwarzschild solution deals with. A given black hole appears to have only 1 state that is unconditionally characterized by its mass M, angular momentum L, and charge Q. With $0 = Q = L = M$, there is no black hole, and Eq. (12) is consistent with $S_{Bek}$ since $A = 0$. With $M \ne 0$, there would appear to be only 1 state making Eq. (12) = 0, which is inconsistent if not totally incompatible with $S_{Bek}$ which is $\propto A \ne 0$.

The interpretation of this paper with N constituents inside the black hole avoids this inconsistency. Furthermore it permits straightforward exploration of the dynamics and time scale by which equilibrium is achieved in terms of the internal collision frequency of the constituents. Though black holes are far from being well understood, the general orthodox view is that black hole entropy as related to the black hole's area is on a firm and well understood basis.  This aspect of black holes has been most diffficult to fathom because it is contrary to all other known aspects of the entropy principle where entropy is an extensive quantity and the volume of a system makes a major contribution to its entropy.  We still have much to learn about black holes as they are clandestine about their interiors, hiding a theoretical gravitational singularity, but most likely not a real physical singularity.

Not only has an equivalent volume contribution for black hole entropy been found , but we also found out something about the number of constituents.  In a recent extensive review, Bekenstein (2003) examines the relationship between black hole entropy and information theory.  However, the fundamental tenet that black hole entropy is only a function of its area is not questioned.  Two biases have contributed to the view that black hole entropy is purely related to the area of a black hole.  One is that we can have no knowledge of what is inside a black hole.  On the basis of my analysis, I would modify this statement to "no direct knowledge."  The other is an historical bias since in their paper on the laws of thermodynamics of black holes, Bardeen, Carter, and Hawking (1973) showed that the increase of the area of a black hole is analogous to entropy.

Black hole temperature is obtained quantum mechanically by means of an analysis (Hawking, 1974, 1975) that is unconventional with respect to thermodynamics and statistical mechanics.  This makes temperature seem like a not well-defined concept for a black hole, without consideration of the constituents introduced in this paper.  Temperature is fundamentally a statistical concept requiring many bodies.  If the distribution of energies is broad, the temperature is high.  If the distribution of energies is narrow, the temperature is low.  Temperature is basically a measure of the half-width of the distribution. For a black hole it not clear what the distribution is or what the states are for determining the entropy.

In their paper on the laws of thermodynamics of black holes, Bardeen, Carter, and Hawking (1973) very specifically said that the temperature they assigned was not real, but just an effective temperature. They also were very careful to add that for them the real temperature of a black hole is zero. The latter was the basis on which Hawking attacked Bekenstein's concept of black hole entropy. Hawking argued that if a black hole has entropy this implies that a black hole has a real temperature greater than 0, and that we all know that the temperature of a black hole must be 0. If a black hole has a real temperature (is thermalized), then it must radiate leading to *reductio ad aburdum*; but Hawking made this his argument before discovering Hawking radiation.

## 6. CONCLUSION

Black hole density led to the simplest possible equation for black hole radiation. Similarly black hole density led to a novel insight into black hole entropy. Black hole entropy may prove to be as malleable as the conservation of energy. Whenever energy conservation was challenged, a way was found to preserve it. It is not clear why my simple model works so well. For little black holes this may be due to asymptotic freedom because the constituents can be at extremely high density. Another possibility is that there are neutralizing effects similar to those that permit a free gas model of electrons inside a metal. Nevertheless this model addresses and to some degree answers the question of why the entropy of a black hole is directly related to its horizon area, even in the non-equilibrium state. It says that the black hole area reflects the number and entropy of the constituents inside it.